\documentclass[a4paper]{jpconf}
\usepackage[dvips]{graphicx}
\graphicspath{{figures/}}

\newcommand{\mybibitem}[6]%
	   {\bibitem{#1} #2 #3 {\it #4} {\bf #5} #6}
\newcommand{\collab}[1]{[#1 collaboration]}
\newcommand{\apj}{Astrophys. J.}
\newcommand{\apjl}{Astrophys. J. Lett.}
\newcommand{\apjss}{Astrophys. J. Suppl. Series}
\newcommand{\prl}{Phys. Rev. Lett.}
\newcommand{\prd}{Phys. Rev. {\rm D}}
\newcommand{\nat}{Nature}
\newcommand{\astrpart}{Astropart. Phys.}
\newcommand{\asas}{Astron. Astrophys.}
\newcommand{\mnras}{Mon. Not. R. Astron. Soc.}
\newcommand{\asr}{Advan. Space Reaserch}
\newcommand{\rpp}{Rep. Prog. Phys.}
\newcommand{\pr}{Phys. Rep}
\newcommand{\acp}{AIP Conf. Proc.}
\newcommand{\jcap}{Journ. Cosm. Astropar. Phys.}
\newcommand{\sas}{Soviet Astr.}

\begin{document}
\title{Fundamental physics in space with the Fermi Gamma-ray Space Telescope}

\author{Luca Baldini (for the Fermi LAT Collaboration)}

\address{INFN-Sezione di Pisa, Largo B. Pontecorvo 3, 56127, Pisa, Italy}

\ead{luca.baldini@pi.infn.it}

\begin{abstract}
  Successfully launched in June 2008, the Fermi Gamma-ray Space Telescope,
  formerly named GLAST, has been observing the high-energy gamma-ray sky
  with unprecedented sensitivity for more than two years, opening a new
  window on a wide variety of exotic astrophysical objects.
  This paper is a short overview of the main science highlights, aimed at
  non-specialists, with emphasis on those which are more directly connected
  with the study of fundamental physics---particularly the search for signals
  of new physics in the diffuse gamma-ray emission and in the cosmic radiation
  and the study of Gamma-Ray Burst as laboratories for testing possible
  violations of the Lorentz invariance.
\end{abstract}

\section{Introduction\label{sec:intro}}

Designed to survey the gamma-ray sky in the broad energy range from 20~MeV to
more than 300~GeV, with the additional capability of studying transient
phenomena at lower energies, Fermi is \emph{de facto} the reference
space-based gamma-ray observatory of this decade. It largely surpasses the
previous generations of gamma-ray telescopes in terms of effective area,
energy range, instrumental dead time, angular resolution and field of view.
It has the ability to observe 20\% of the sky at any time which, in the
nominal scanning mode of operation, allows to view the entire sky every three
hours.

The Fermi observatory carries two instruments on-board: the Gamma-ray Burst
Monitor (GBM)~\cite{gbm_paper} and the Large Area Telescope
(LAT)~\cite{lat_paper}.
The GBM, sensitive in the energy range between 8~keV and 40~MeV, is designed
to observe the full sky not occulted by the Earth with rough directional
capabilities (at the level of one to a few degrees) for transient point
sources, particularly Gamma-Ray Bursts (GRBs).

The LAT is a pair conversion telescope for photons above 20~MeV (up to
hundreds of GeV). Though owing most of the basic design to its
predecessors---particularly the Energetic Gamma Ray Experiment Telescope
(EGRET)~\cite{egret_paper} on-board the CGRO mission---it exploits the state
of the art in terms of detector technology, which allows for a leap in
sensitivity of a factor of ten or more.  The design, construction and
operation of such a complex detector is a fascinating subject on its own and
the interested readers can refer to~\cite{lat_paper} and references therein. 

Based on the talk that I presented at the fifth international workshop DICE
(held in Castiglioncello, Italy, in September 2010), this is a very
\emph{biased} and unconventional review of the Fermi science highlights of
the first two years of operation. In fact a significant part of the material
that would naturally belong to a review paper is deliberately omitted, here,
in favor of a few items of discussion that I think will be of interest for
the conference audience. For a more orthodox and comprehensive overview, the
reader can refer to~\cite{lat_1year_highlights} and references therein.

\subsection{The gamma-ray sky\label{subsec:gamma_sky}}

\begin{figure}[!htbp]
\includegraphics[width=0.7\textwidth]{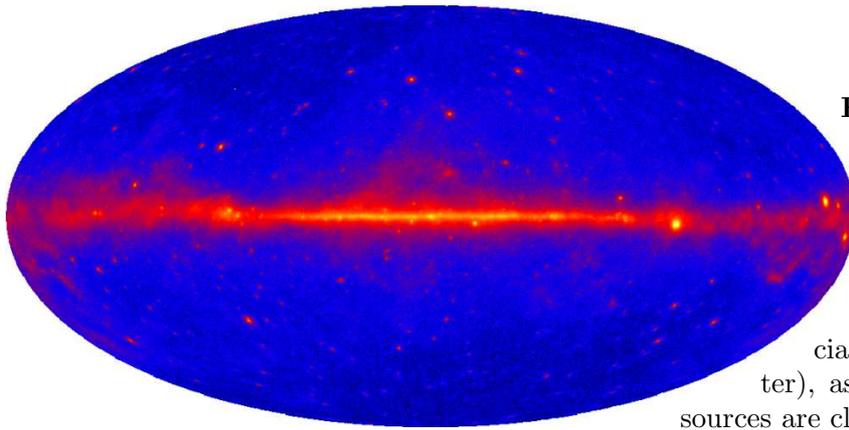}%
\hspace*{-0.145\textwidth}%
\begin{minipage}[b]{0.455\textwidth}%
  \caption{\label{fig:skymap}
    \parshape=10
    60pt 140pt
    72pt 128pt
    78pt 122pt
    82pt 118pt
    80pt 120pt
    75pt 125pt
    68pt 132pt
    50pt 150pt
    30pt 170pt
    0pt 200pt
    Sky map of the gamma-ray candidates above 200~MeV collected by the LAT in
    the first year of operation. The diffuse emission along the Galactic plane
    (and especially around the Galactic center), as well as the brightest
    point sources are clearly visible.}
\end{minipage}
\end{figure}

Figure \ref{fig:skymap} shows the full sky above 200~MeV as seen by the Large
Area Telescope; it is a simple map in Galactic coordinates of the event rate
corrected for the exposure, with no further processing. The most prominent
feature is the plane of our Galaxy, which is indeed very bright in gamma-rays. 

Quite a few point sources can be seen, both along the Galactic plane and at
high celestial latitudes (most of the latter being of extragalactic origin).
In fact a detailed analysis of the LAT data reveals many more than
those evident at first glance: the first Fermi LAT catalog~\cite{lat_1fgl},
based on one year's worth of data, contains 1451 sources---to be compared with
the 271 detected by the EGRET telescope over its entire mission
lifetime~\cite{egret_3eg}. Though one of the main science objectives of the
Fermi mission, the study of gamma-ray point sources is beyond the scope of
this review.

The vast majority of what remains in the sky-map once the contribution from
point sources is subtracted is produced by \emph{diffuse} processes taking
place in our Galaxy---primarily the interaction of charged Cosmic
Rays~(CR) with the Galactic interstellar gas and radiation fields. This
component (concentrated onto the Galactic plane but not limited to it) goes
under the name of Diffuse Galactic Emission~(DGE) and accounts for some
90\% of the total gamma-ray intensity in the LAT energy range.

The presence of a much fainter isotropic emission, commonly known as the
Extragalactic Gamma-ray Background (EGB) is a firmly established experimental
fact~\cite{lat_egb}. The search for a possible truly diffuse component
(on top of contributions from unresolved faint sources and misidentified CR
background) is one of the primary scientific interests associated to this type
of radiation.
The DGE and EGB and their relevance in the search for hints of new physics
will be discussed in more detail in section \ref{sec:diffuse}.

\section{Search for photon lines\label{sec:lines}}

The quest for a possible narrow line in the diffuse gamma-ray background
arises naturally in the context of indirect Dark Matter (DM) searches,
where one can look for photons produced in two-body DM particle annihilations
$\chi\chi \rightarrow \gamma X$ or decays $\chi \rightarrow \gamma X$.
Since in most scenarios dark matter is electrically neutral (and therefore
does not couple directly to photons) such processes only occur at higher
orders and the corresponding expected branching ratios are strongly suppressed.
Despite this, a photon line, if present, is comparatively easy to identify
and distinguish from the standard astrophysical sources of gamma rays---whose
flux is dominant in most situations. Unlike most of the cases that we shall
examine in the following sections, this channel is peculiar in that it features
a distinctive experimental signature that, if observed, would incontrovertibly
indicate new physics at work.

\begin{figure}[htbp]
  \includegraphics[width=0.55\textwidth,trim=0 0 0 50pt]%
                  {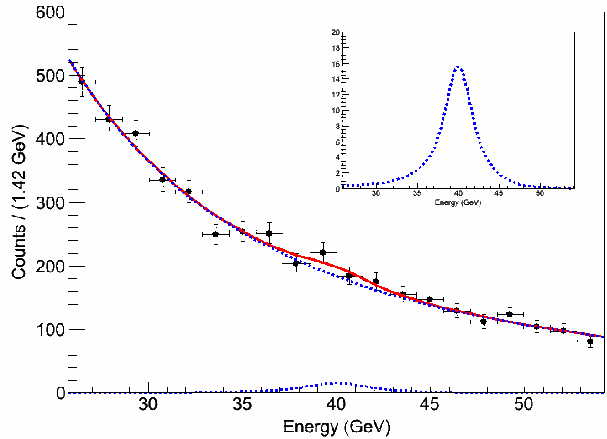}%
  \begin{minipage}[b]{0.45\textwidth}%
    \caption{\label{fig:line}
    Binned representation of the fit procedure (here centered at 40~GeV)
    used to derive the upper limit on the flux of a possible photon line
    contribution in the all-sky (except for part of the Galactic plane)
    spectrum (from \cite{lat_photon_lines}).
    The two dotted lines represent the background (modeled with a
    power law) and the signal from the fit, respectively, while the red line
    is their sum. The insert shows a close-up of the instrument response
    to a monochromatic line at 40~GeV, which is used to model the signal.}
  \end{minipage}
\end{figure}

Sure enough, the detector response to a monochromatic line is \emph{not} a
monochromatic line and the effect of the finite energy resolution can hardly
be ignored. That said, this response can be modeled by means of Monte Carlo
simulations (see the insert in Figure \ref{fig:line}) and verified with tests
at accelerators so that it can be effectively folded into the procedure used
to asses the statistical significance of a possible line component in the
measured count spectrum.

No significant evidence of gamma-ray line(s) has been found, based on the
first 11 months of data, between 30 and 200 GeV~\cite{lat_photon_lines} and
work is ongoing to extend the energy range of the analysis and include more
data. The detailed discussion of the upper limits obtained and their relevance
in the context of specific models is beyond the scope of this brief overview.

\section{Cosmic rays and diffuse gamma-ray radiation\label{sec:diffuse}}

Galactic cosmic rays and gamma rays are closely related to each other. As
mentioned in section~\ref{subsec:gamma_sky}, the diffuse gamma-ray emission
is mainly produced by cosmic-ray protons and He nuclei interacting with the
interstellar gas (through subsequent $\pi^0$ decay) and by Cosmic-Ray Electrons
(CRE) interacting with the gas itself (via bremsstrahlung), with the
ambient low-energy photons (via inverse Compton scattering) and with the
Galactic magnetic field (via synchrotron emission). As a
consequence, there is a conspicuous number of different ingredients going into
the modeling of the Galactic diffuse emission: the energy spectra of the
various cosmic-ray species, the interaction cross sections, the distribution
of the atomic and molecular gas in the interstellar medium and the details of
the interstellar radiation field~\cite{cr_review}.
Most of these ingredients, indeed, are known
to a sufficient level of accuracy to allow realistic numerical simulations of
the high-energy Galaxy in which the CR and gamma-ray observables can be
self-consistently modeled (see for instance~\cite{galprop}). Still, there are
significant uncertainties connected with the predictions of those models. And
it goes without saying that such uncertainties are, by their nature, of primary
importance to any search for signatures of new physics.

Over the last few years there has been a tremendous interest in the possibility
of exotic sources of cosmic rays and/or gamma rays. This excitement has been
indeed fostered by a number of measurements supposedly indicating unexpected
\emph{features} in the basic observables.
Terms like ``GeV excess'', ``ATIC peak'' or ``PAMELA positron fraction''
(which we shall discuss in detail in a moment) have become part of the common
jargon of the community, with a profusion of scientific papers aimed at
interpreting the apparent ``anomalies'' and the ``excesses'' in the data.

Overall, if we look back at the beginning of 2009 (i.e. a few months after the
launch of Fermi), it is fair to say that several independent pieces of
experimental evidence were (to different extents) difficult to reconcile
with our basic understanding of the cosmic-ray propagation in the Galaxy.
As we shall see in this section, the situation has largely changed over the
last two years.

\subsection{Galactic diffuse radiation}

The Galactic diffuse gamma-ray emission is not only important in itself
for the strong scientific interest it bears; it is also important as it
constitutes a bright foreground for the analysis of point sources
(influencing the determination of their position and flux) and for the
much fainter isotropic diffuse component.

\begin{figure}[!hbtp]
  \begin{minipage}[t]{0.48\textwidth}
    \includegraphics[width=\textwidth,trim=0 20pt 0 10pt]%
                    {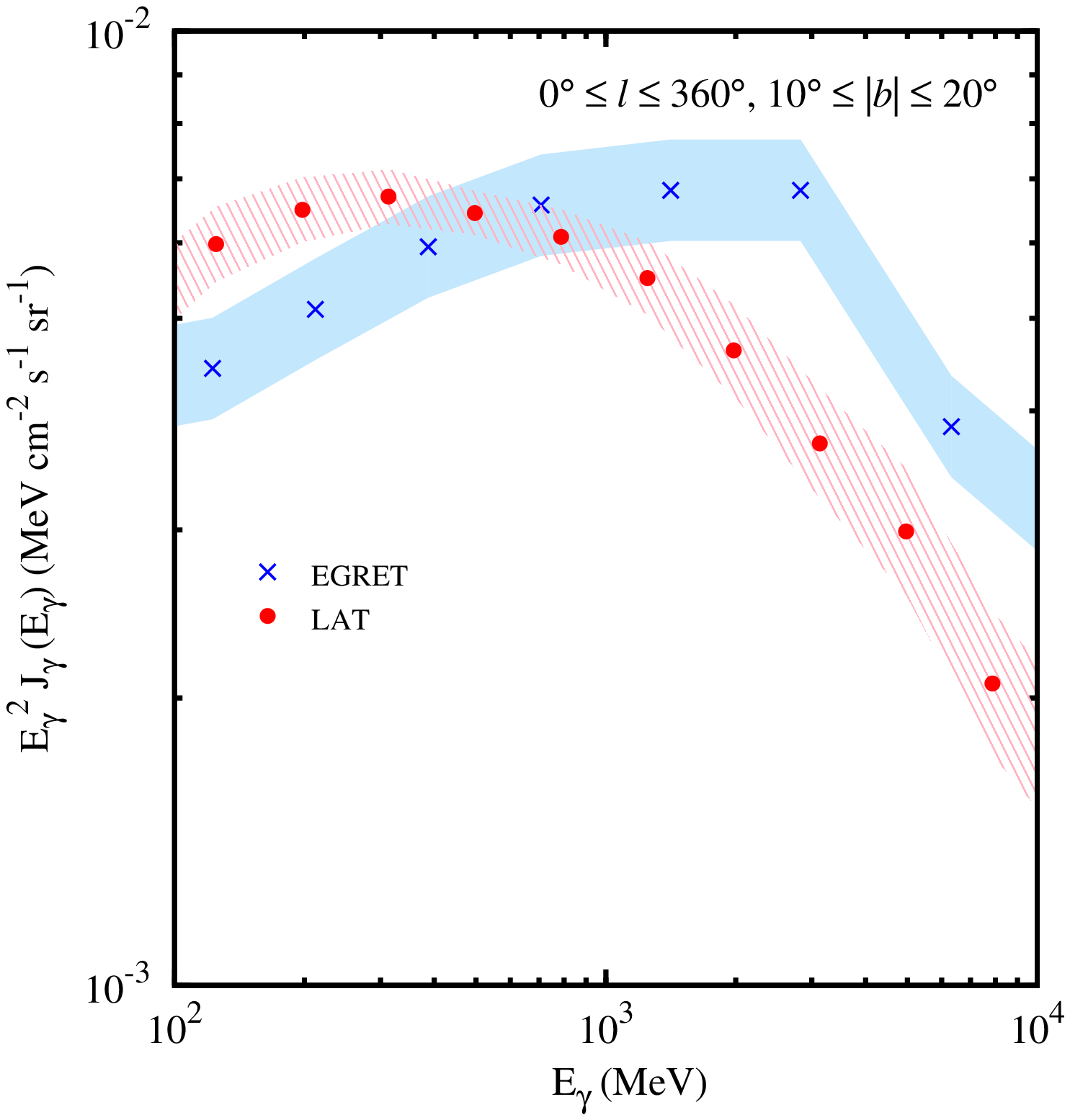}
    \caption{\label{fig:dge_egret}
      Average diffuse gamma-ray emission intensity at intermediate galactic
      latitudes $10^\circ \le \left| b \right| \le 20^\circ$ (from
      \cite{lat_gal_diffuse}). The shaded/hatched bands superimposed to the
      data points represent the systematic uncertainties for the Fermi LAT
      (red) and the EGRET (blue) instruments, respectively. The EGRET
      ``GeV excess'' is not confirmed by the LAT.}
  \end{minipage}\hfill%
  \begin{minipage}[t]{0.48\textwidth}
    \includegraphics[width=\textwidth,trim=0 20pt 0 10pt]%
                    {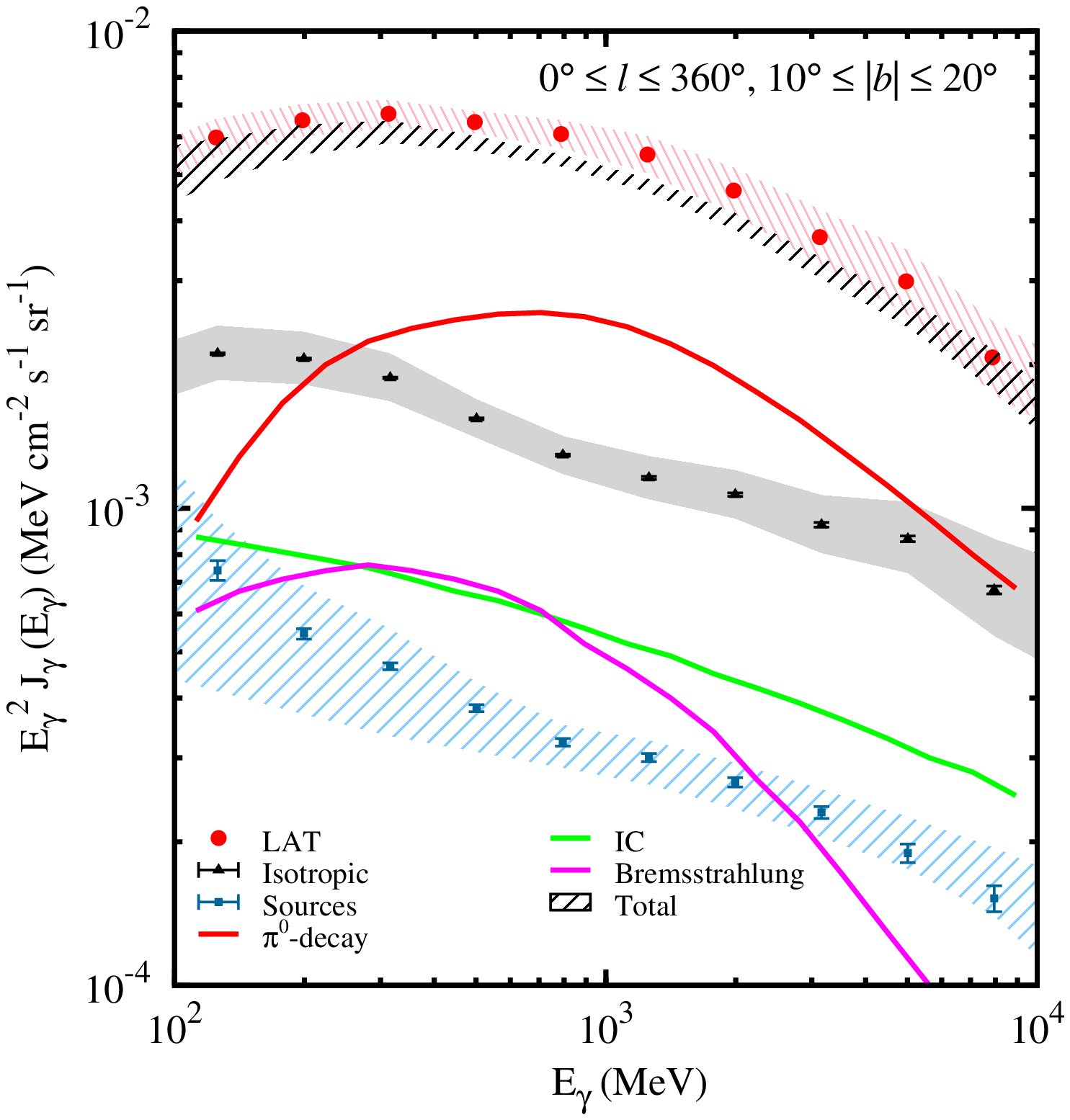}
    \caption{\label{fig:dge_model}
      Comparison of the LAT spectrum shown in figure \ref{fig:dge_egret}
      with an \emph{a priori} diffuse Galactic emission model (black hatched
      band). The model components ($\pi^0$ decay, Inverse Compton~(IC) and
      bremsstrahlung) are included for completeness, along with the
      contributions of point sources and isotropic background,
      derived from the fit to the LAT data.}
  \end{minipage} 
\end{figure}

The Fermi LAT collaboration has measured the spectrum of the DGE at
intermediate latitudes, between 100~MeV and 10~GeV, based on the first 5~months
of operations~\cite{lat_gal_diffuse}. The spectrum is significantly softer than
that measured by EGRET (Figure~\ref{fig:dge_egret}) and it is consistent with
an a-priori---i.e. reflecting the state of the art in terms of local cosmic-ray
data but not tuned to reproduce the gamma-ray data---model
(Figure~\ref{fig:dge_model}). Particularly, the excess emission above 1~GeV
measured by EGRET, which in the literature is often informally referred to
simply as the ``GeV excess'', is not confirmed by the LAT.

The region of interest $10^\circ \le \left| b \right| \le 20^\circ$ (or about
$15^\circ$ from the Galactic plane) was chosen to maximize the signal to noise
ratio and, at the same time, minimize the uncertainties in the modeling
(in order to model the diffuse emission on the Galactic plane it is necessary
to trace the cosmic rays through the whole Galaxy). Work is ongoing in the
Collaboration for a more thorough analysis, not limited to the region of
interest and the energy range in~\cite{lat_gal_diffuse}.
It is already clear, however, that most of the scientific exegesis of the
original EGRET claim is to be radically reviewed in the light of the new Fermi
results.

\subsection{Isotropic diffuse radiation}

First detected by the second Small Astronomy Satellite (SAS-2)
mission~\cite{sas2_egb} and later confirmed by EGRET~\cite{egret_egb}, the
extragalactic diffuse gamma-ray background is generally considered to be
the superposition of contributions from unresolved extragalactic sources and
truly diffuse emission processes---the latter potentially including
imprints of new physics.

The measurement of the EGB is, to many respects, more challenging than the
others being reviewed in this paper. Not only because it is intrinsically
faint, but also because, being isotropic and stationary, it is not possible to
exploit any spatial and/or temporal signature to enhance the signal-to-noise
ratio. In fact the estimation of the residual contamination of misidentified
charged cosmic rays (whose flux is also isotropic) is one of the critical
aspects of the analysis.

\begin{figure}[!htbp]
  \centering\includegraphics[width=0.7\textwidth]{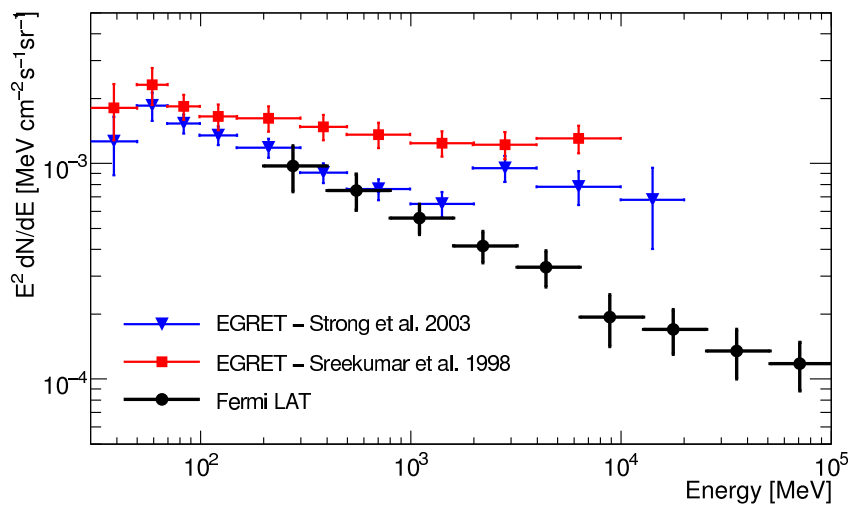}\hfill%

  \caption{\label{fig:egb_sed}
    EGB intensity measured by Fermi (from~\cite{lat_gal_diffuse}), compared
    with the one derived from the EGRET data in~\cite{egret_egb} and the
    re-analysis~\cite{egret_egb_new}, based on a different model for the
    Galactic diffuse emission.}
\end{figure}

Some 5~years after the original EGRET publication~\cite{egret_egb}, a
re-analysis of the EGRET data~\cite{egret_egb_new}, based on a different
model of the diffuse Galactic emission, revealed a distinctive peak in the
spectrum around 3~GeV, which triggered the interest of the community in what
might have been its origin. To this respect, it is interesting to note
how the blue and red data points in figure~\ref{fig:egb_sed} are based on
the \emph{same} experimental data. The substantial difference between the two
spectra is only attributable to a difference in the underlying model for the
DGE (which, as we mentioned before, is a strong foreground for the measurement
of the EGB). This, in turn, is a good illustration of how often putative
signs of new physics are inextricably mixed with genuine astrophysical
processes, and the uncertainties on the latter can have significant effects.
For what it's worth, it is the opinion of the author that this aspect does not
always receive the attention it deserves. 

The spectrum of the extragalactic gamma-ray background measured by the
Fermi~LAT~\cite{lat_egb}, shown in Figure~\ref{fig:egb_sed}, is significantly
softer than the EGRET spectrum. It is compatible with a featureless power law
with spectral index $\Gamma = 2.37 \pm 0.05$ and does not confirm the spectral
feature evident in the EGRET re-analysis. A detailed study of the Fermi source
count distributions~\cite{lat_source_count_egb} shows that unidentified
Blazars can only account for 40\% of the EGB flux at most. The nature of the
remaining fraction is of the highest scientific interest and will be
investigated as more data are collected.

\subsection{Leptons in cosmic rays}

Charged cosmic rays have been extensively studied since the mid 1960s with
balloon-borne and satellite experiments---and, at comparatively higher
energies, by means of ground arrays. We have, to date, a large,
well-established body of knowledge which includes spectra, relative
abundances, composition and time variation. Among the different species,
energetic cosmic-ray electrons and positrons are peculiar in that they
rapidly loose energy through synchrotron radiation in the Galactic magnetic
field and inverse Compton scattering on the interstellar radiation field.
Therefore, at high energy, they effectively probe the nearby Galactic space.

\begin{figure}[!htbp]
  \centering\includegraphics[width=0.80\textwidth]{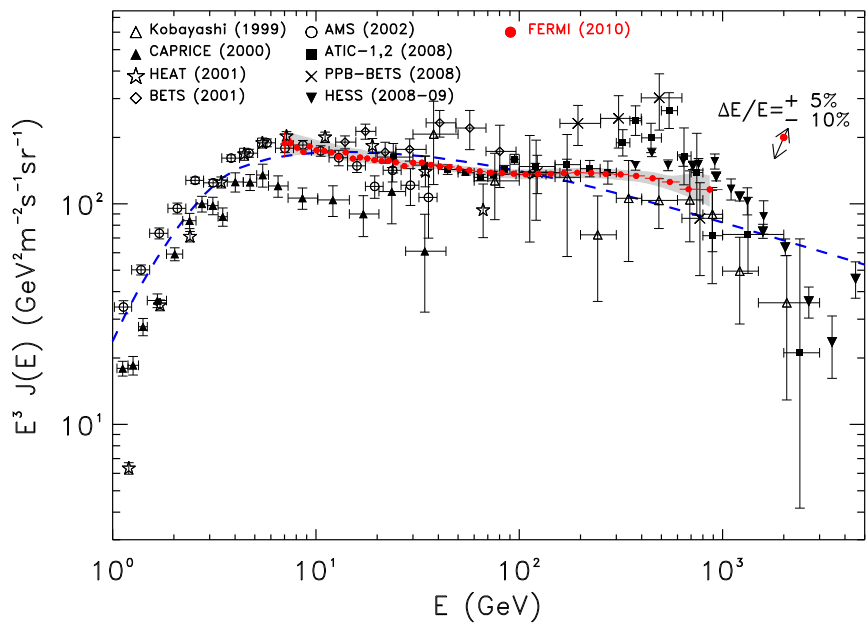}

  \caption{\label{fig:cre_spec}
    Cosmic-ray electron spectrum measured by the Fermi LAT (red points, from
    \cite{lat_cre_full}). The gray shaded band indicates the systematic
    uncertainties associated with the flux values, while the blue dashed line
    represents the prediction of a diffusive propagation model tuned on the
    pre-Fermi data~\cite{pre_fermi_diff_model}.
    The other recent measurements included are from
    \cite{kobayashi_cre, caprice_cre, heat_cre, bets_cre, ams_cre,
      atic_cre, ppbbets_cre, hess_cre_high, hess_cre_low}.}
\end{figure}

In 2008 the balloon-borne Advanced Thin Ionization Calorimeter (ATIC)
experiment reported evidence for a prominent, sharp spectral feature between
300 and 800~GeV in the inclusive $e^+ + e^-$ spectrum~\cite{atic_cre}---which
has become widely known as the ``ATIC peak''. One of the possible
interpretations put forward by the authors was the existence of a Kaluza-Klein
particle at the TeV scale annihilating directly into electron-positron pairs.
This measurement, selected by the American Institute of Physics as one of the
10 Top Physics Stories of 2008, has stirred tremendous interest in the
high-energy astrophysics community, with the ATIC peak often discussed in
conjunction with another measurement, published shortly after by the
PAMELA collaboration---namely the evidence that the fraction of positron in
the $e^+ + e^-$ spectrum rises with energy~\cite{pamela_pos}.

In 2009 the Fermi LAT collaboration has published the first systematic-limited
measurement of the spectrum of cosmic-ray electrons and
positrons~\cite{lat_cre} between 20~GeV and 1~TeV. Though harder than
previously thought, the Fermi spectrum does not show evidence of any prominent
feature. This has been later confirmed by the H.E.S.S.
experiment~\cite{hess_cre_low}, as well as by a dedicated analysis of the
Fermi data aimed at the highest end of the energy range~\cite{lat_cre_full}.

Figure~\ref{fig:cre_spec} shows the Fermi LAT spectrum, compared with the
predictions of \emph{a} pre-Fermi model~\cite{pre_fermi_diff_model}.
Whether the difference between the two is due to the limits in our
understanding of the cosmic-ray propagation in the Galaxy or to the presence
of additional (astrophysical or exotic) sources of high-energy electrons is a
question with no definitive answer, at this point~\cite{grasso_cre_int}.

\subsection{Further comments and remarks}

The results from the first two years of operation of the Fermi LAT have
radically changed our understanding of the diffuse gamma-ray radiation and 
cosmic-ray propagation. Most of the claims of prominent departures from the 
accepted scientific paradigms, in fact, have not been confirmed and might
be, at least partially, just instrumental artifacts.
It is probably fair to say that the positron fraction measured by the PAMELA
experiment is still at odds with the standard picture of secondary cosmic-ray
production. The Fermi LAT and the forthcoming Alpha Magnetic Spectrometer
(AMS-02) space experiment will help to shed light on the
subject~\cite{ams02_pos}.
In the meantime papers like~\cite{blasi_cre} and~\cite{waxman_cre} might
be telling us that what we really learn from the ``lepton excesses'' is that
we don't know enough astrophysics, yet.

That all said, the overall tone of this section should not be perceived as too
dismissive. Diffuse gamma-ray emission and cosmic-rays, when interpreted as
different pieces of a single puzzle, still constitute a very exciting
perspective for the quest of new physics in space---particularly in the
context of indirect dark matter searches. Along with the energy spectra
additional observables (i.e. the anisotropies in the arrival
directions~\cite{lat_cre_any}) will be of great interest, as new, more
accurate measurements become available and our understanding of the
astrophysical environment improves.

\section{Gamma-Ray Bursts and tests of Lorentz invariance
  violation\label{sec:grb}}

Gamma-Ray Burst (GRB) are the most violent explosions in the Universe. The
isotropic energy emission, that can be estimated based on the measured fluence,
can be as high as $10^{54}$~erg for the brightest bursts, with a huge energy
release in keV--MeV gamma-rays. This naturally suggests a narrow beaming of
the emission---still, a beaming factor of the order of $10^{-3}$ implies a
considerable energy budget of $10^{51}$~erg.

\begin{figure}[!htbp]
  \centering\includegraphics[width=0.87\textwidth]{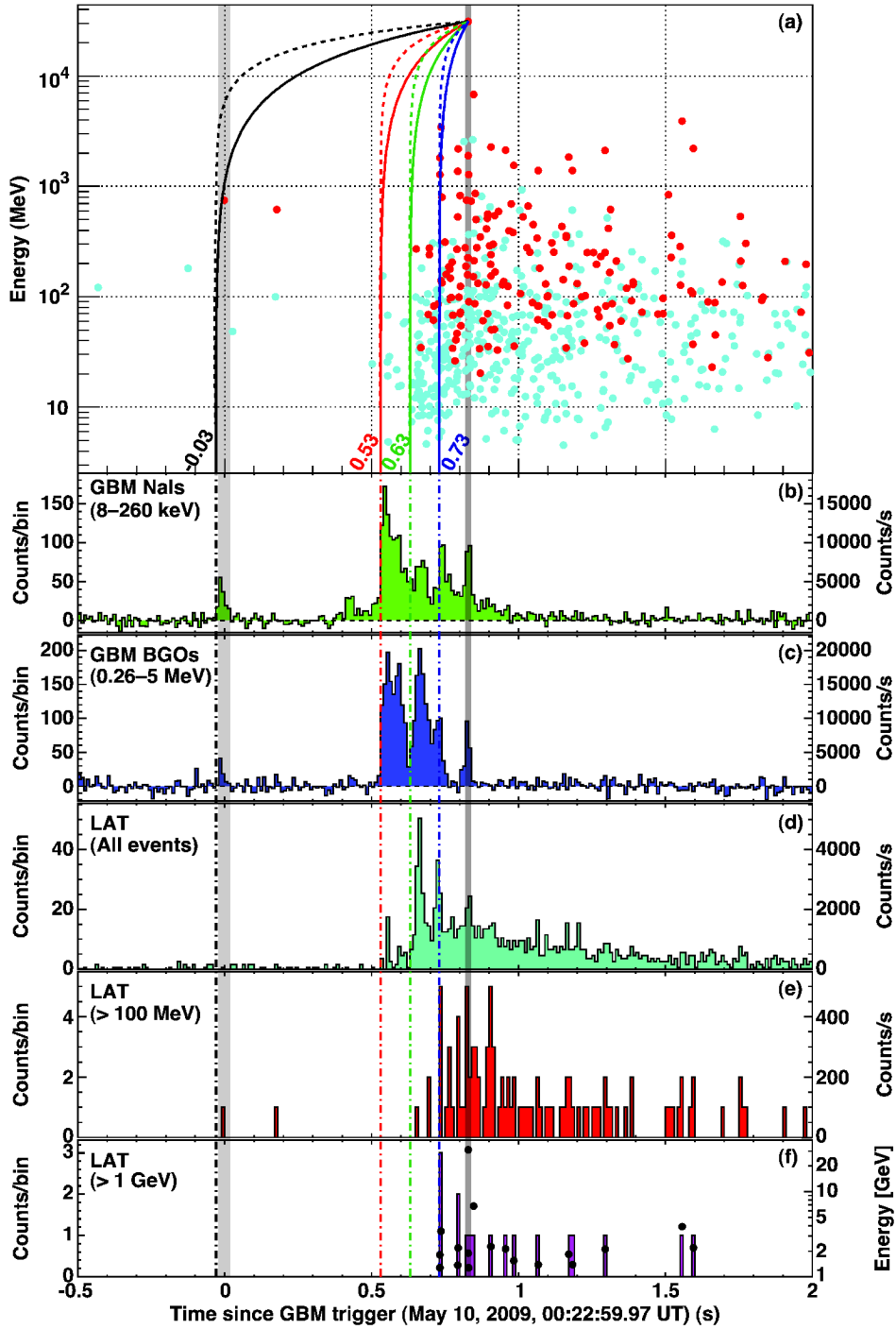}

  \caption{\label{fig:grb090510}
    Light curves of the short GRB 090510 in different energy bands
    (from \cite{lat_liv}). The top panel shows the photon energy vs. the photon
    arrival time for two different event selections (see \cite{lat_liv} for
    more details). The remaining plots are the GBM and LAT light curves from
    lowest to highest energy (top to bottom). The overlaid curves, which are
    normalized to pass through the highest energy (31~GeV) photon, represent
    the relation between the energy and the arrival time for linear
    (solid line) and quadratic (dashed line) LIV---assuming that the photon is
    emitted in correspondence of the GBM onset (30~ms before the trigger
    time).}
\end{figure}

GRBs have been extensively studied at relatively low energies over the last
two decades, the most notable observed features being:
\begin{itemize}
\item the cosmological origin (they are isotropically distributed in the sky
  and have been observed up to a redshift $z \approx 8$);
\item the bimodal duration distribution (with \emph{short} bursts lasting
  for $\approx 1$~s and \emph{long} ones lasting for tens of s);
\item the rapid variability of the light curves, down to time scales of
  the order of the ms.
\end{itemize}
The Fermi LAT has effectively opened a new observational window, enabling the
systematic study of the prompt emission above 100~MeV on a large sample of
bursts---with supporting observations by the GBM at lower energy being crucial
in putting the new information in the context of what is already known
(see Figure \ref{fig:grb090510}).
The GBM detects about 250~GRB per year, roughly half of them being in the LAT
field of view; some 10\% of the latter (or $\approx 10$ per year) are bright
enough at high energy to be detected by the Large Area Telescope.
As of the end of 2010, the Fermi LAT collaboration has reported the detection
of 22 GRBs through the Gamma-ray Burst Coordination Network (GCN)~\cite{gcn}.

From the standpoint of this brief overview, gamma-ray bursts are mainly
interesting as laboratories to test possible violations of the Lorentz
Invariance (LIV). There are indeed several theoretical
frameworks---particularly in the context of some of the Quantum Gravity (QG)
formalisms currently being investigated---that predict (or can accommodate) a
modification of the standard photon dispersion
relation~\cite{liv_1, liv_2, liv_3}. This happens in the form
of an energy-dependent term ceasing to be negligible at some mass scale
$m_{\rm QG}$, possibly of the order of the Planck mass
$m_{\rm P} \approx  1.22 \times 10^{19}$~GeV:
\begin{equation}
E^2 = p^2 c^2 + \Delta_{\rm QG} (E, p^2, m_{\rm QG})
\end{equation}
One can therefore imagine to expand the additional term in series of
$E/m_{\rm QG}$ up to the first non-vanishing order $n$, which yields a non
trivial expression for the speed of light in vacuum as a function of the
photon energy:
\begin{equation}
v(E) = c \left[1 \pm \left( E/m_{\rm QG}c^2 \right)^n \right]
\end{equation}
where the $\pm$ sign distinguishes the \emph{subluminal} and the
\emph{superluminal} cases. The experimental signature of such a scenario would
be the fact that two photons of different energies, emitted simultaneously,
will travel with different velocities and arrive to the observer with some
relative time delay. Even a tiny variation in the speed of light, when
accumulated over cosmological distances, can in principle be revealed at
energies much lower than the characteristics scale $m_{\rm QG}c^2$. If, for the
sake of simplicity, we limit ourselves to linear LIV (i.e. $n = 1$) the
relevant expression for the time delay of two photons emitted by a distant
astrophysical source at redshift $z$ is:
\begin{equation}
\Delta t = \pm \frac{1}{H_0} \frac{\Delta E}{m_{\rm QG}}
\int_0^z \frac{(1 + \xi)}{\sqrt{\Omega_\Lambda + \Omega_m (1 + \xi)^3}} d\xi
\end{equation}

The short duration, rapid variability and cosmological distance make GRBs
for perfect candidates to constraint possible Lorentz invariance violations.
Though the Fermi collaboration has previously set stringent lower limits on
the mass scale $m_{\rm QG}$ with the long GRB~080916C~\cite{lat_grb080916c},
here we shall limit the discussion to the short GRB~090510~\cite{lat_liv}.
As shown in Figure~\ref{fig:grb090510}, this burst features a 31~GeV photon
emitted 0.829~s after the GBM trigger time. Thanks to the synergy with other
x-ray and optical instruments, it was possible to measure the redshift of the
source, which turned out to be $z = 0.903 \pm 0.003$. Sure enough, we don't
know \emph{exactly} when this particular photon was emitted, relative to the
much more abundant low-energy gamma-rays, so that this arrival time cannot be
readily interpreted as a \emph{time delay}. However, if we assume that the
photon itself was not emitted before the beginning of the precursor of the
burst (30~ms before the GBM trigger), that gives us all the necessary
ingredients to set a robust lower limit $m_{\rm QG} > 1.19 m_{\rm P}$ in the
linear ($n = 1$) subluminal case. In fact this is the most stringent limit of
its kind available so far, and, effectively, it strongly disfavors any
scenario characterized by a linear, subluminal LIV. The assumption on which
it is based is very reasonable and somehow supported by the fact that in
none of the bursts detected by the Fermi LAT high-energy photons were detected
before the onset of the low-energy emission---on the contrary, in many cases
there was evidence for a spectral evolution with a consequent delay of the
high-energy emission onset. By using more sophisticated approaches (such as
the one described in~\cite{discan}) or relaxing the basic assumption about the
emission time of the highest energy photon it is possible to obtain
comparable or more stringent limits, in both the subluminal and the
superluminal case; the reader is referred to~\cite{lat_liv} for further
details.

It is worth stressing that this is another case in which the effect we are
trying to measure (in this case the delay accumulated by the photon during the
propagation) \emph{competes} with another physical effect (i.e. the intrinsic
time delay at the source) which is of the same order or magnitude or bigger
and which is essentially impossible to infer with enough accuracy.
A definitive evidence for a Lorentz invariance violation might come in the
form of the observation of a redshift-dependent effect on a large sample of
bursts, which Fermi might be able to achieve before the end of the mission.

\section{More discussion items and future prospects\label{sec:prospects}}

We conclude this overview by briefly introducing a few more items of
interest, some of which might be the object of future Fermi observations.

\subsection{Pulsar timing array detection of gravitational waves}

Pulsars constitute one of the primary scientific targets for Fermi---in fact
the Large Area Telescope has already increased the number of known gamma-ray
pulsars by roughly an order of magnitude from the 6 detected by EGRET.
Though most of the scientific interest associated with these discoveries is
of purely astrophysical nature, Milli-Second Pulsars (MSP)---old neutron stars
spinning at a rate of hundreds of revolutions per seconds---are effectively
accurate cosmic clocks suitable for tests of fundamental physics.

It was suggested already in the late 1970s~\cite{pta_1, pta_2} that
gravitational waves could induce small perturbations in the arrival times of
the radio pulses and that those perturbations could be potentially
detectable. It actually did not take too long to realize that monitoring the
time residuals of an \emph{array} of pulsars had the potential for a
significant increase in sensitivity~\cite{pta_3}. Based on this idea,
\emph{Pulsar timing arrays} are meant to observe, over periods of years, the
small, correlated effects that gravitational waves have on the arrival times
of the radio pulses at Earth. The whole subject recently experienced a renewed
interest as, by studying unassociated high-energy sources detected by the Fermi
LAT in the first months of operation, radio astronomers discovered 17 new
millisecond pulsars (corresponding to a 30\% increase in the overall
number of known MSPs). This is really the product of a synergic cooperation
between the Fermi Collaboration and the largest radio telescopes in the world:
in many cases the gamma-ray data are too sparse to detect pulsation in the
unidentified sources, whereas radio astronomers can take advantage of the
positions of potential pulsar candidates to complement time-consuming sky
surveys and improve the discovery reach.

Pulsar timing already provides the most compelling experimental evidence for
the existence of gravitational radiation through the orbital decay of the
binary system PSR~1913+16~\cite{psr1913+16}. However, this is generally
regarded as an \emph{indirect} evidence, in that what is observed is the
effect of the emission on the emitter. On the other hand, pulsar timing arrays
really act as gravitational wave detectors. As more millisecond pulsars are
identified (and monitored by radio telescopes), they could provide the
first \emph{direct} evidence for the existence of gravitational radiation in
the ultra low-frequency regime, where the ground based interferometers are not
sensitive.

\subsection{Gravitational lensing}

The light curves of distant, gravitationally lensed Active Galactic Nuclei
(AGN) measured by the Large Area Telescope can potentially be used to test
general relativity. Multiple images of strongly lensed sources can be resolved
in the optical band. The study of the time lag between such images (which can
be as large as weeks, due to the different propagation times along the
different light paths) is a standard observational technique, crucial for the
modeling of this kind of systems. This is not directly possible in gamma-rays,
as gamma-ray telescopes do not feature the necessary angular resolution.
Nonetheless, if a flare is detected from such a source, the high-energy light
curve might show evidence for the gravitational \emph{echoes}. A comparison
between the optical and the gamma-ray light curves can test whether photons
differing by eight orders of magnitude in energy are deflected in the same way,
as general relativity predicts.

\subsection{Search for axion-like particles}

Axion-Like Particles (ALP) are hypothetical particles featuring a direct
photon coupling in the presence of an external magnetic field.
Phenomenologically, this gives origin to a neutrino-like oscillation behavior
between ALPs and photons that can possibly lead to observable effects on the
energy spectra of distant gamma-ray sources~\cite{alp}. The photon-axion
mixing can take place both in the gamma ray-source, if this is strongly
magnetized, and in the magnetic fields encountered along the path to the
observer. In the latter case the propagation over cosmological distances
compensates for the weakness of the intergalactic magnetic fields (which is
believed to be of the order of~nG). The prospect of detecting such
modifications to the spectra of gamma-ray sources is complementary to
ground-based experiments such as CAST~\cite{cast} or PVLAS~\cite{pvlas}.

\section{Conclusions\label{sec:conclusions}}

We briefly reviewed some of the science highlights of the Fermi gamma-ray
space telescope after the first two years in orbit. The main message that
this overview is aimed to convey is that the study of fundamental physics
with space-based gamma-ray astronomy is indeed an exciting opportunity.
Gamma-rays directly probe some of the most violent phenomena in the Universe;
they give access to physical systems in the strong gravitational and/or 
magnetic field regime, along with photon propagation over cosmological
distances---in brief, many of the conditions that would be impossible to
set up in a laboratory on the Earth. On the other hand, life in space is
to many respects more difficult than that in a controlled experimental
environment; the physical effects that one is trying to measure are often
interleaved with genuine astrophysical phenomena which (i) cannot be
controlled and (ii) are extremely hard to understand and model to the needed
degree of accuracy.

Fermi turned two years in orbit on June, 2010, and it is definitely
living up to its expectations in terms of scientific results delivered to the
community. The mission is planned to continue at least three more years
(possibly eight) with many remaining opportunities for discoveries.
All the Fermi gamma-ray data (along with the science analysis software tools
developed and used by the Collaboration) are publicly available
through the Fermi Science Support Center~\cite{fssc}.

\ack
The work reported here is a result of the efforts of all members of the
Fermi LAT and GBM Collaborations.

The $Fermi$ LAT Collaboration acknowledges generous ongoing support from a
number of agencies and institutes that have supported both the development and
the operation of the LAT as well as scientific data analysis.
These include the National Aeronautics and Space Administration and the
Department of Energy in the United States, the Commissariat \`a l'Energie
Atomique and the Centre National de la Recherche Scientifique / Institut
National de Physique Nucl\'eaire et de Physique des Particules in France, the
Agenzia Spaziale Italiana and the Istituto Nazionale di Fisica Nucleare in
Italy, the Ministry of Education, Culture, Sports, Science and Technology
(MEXT), High Energy Accelerator Research Organization (KEK) and Japan
Aerospace Exploration Agency (JAXA) in Japan, and the K.~A.~Wallenberg
Foundation, the Swedish Research Council and the Swedish National Space Board
in Sweden.

Additional support for science analysis during the operations phase is
gratefully acknowledged from the Istituto Nazionale di Astrofisica in Italy.

\end{document}